\documentclass[aps,nofootinbib]{revtex4-1}

\usepackage{graphicx}
\usepackage{epstopdf}
\usepackage{latexsym}
\usepackage{amsmath}
\usepackage{amssymb}
\usepackage{color}
\usepackage{mathrsfs}

\usepackage[center]{subfigure}

\begin{document}

 \newcommand{\bq}{\begin{equation}}
 \newcommand{\eq}{\end{equation}}
 \newcommand{\bqn}{\begin{eqnarray}}
 \newcommand{\eqn}{\end{eqnarray}}
 \newcommand{\nb}{\nonumber}
 \newcommand{\lb}{\label}
 \newcommand{\be}{\begin{equation}}
\newcommand{\en}{\end{equation}}
\newcommand{\PRL}{Phys. Rev. Lett.}
\newcommand{\PL}{Phys. Lett.}
\newcommand{\PR}{Phys. Rev.}
\newcommand{\CQG}{Class. Quantum Grav.}

\title{Axial Gravitational Normal Modes of Uniform Density Star in Anti-de Sitter Spacetime}

\author{Kai Lin}\email{lk314159@hotmail.com}
\affiliation{Universidade Federal de Campina Grande, Campina Grande, PB 58429-900, Brasil}
\affiliation{Instituto de F\'isica, Universidade de S\~ao Paulo, S\~ao Paulo, Brazil}

\author{Alan B. Pavan}\email{alan@unifei.edu.br}
\affiliation{Universidade Federal de Itajub\'a, Instituto de F\'isica e Qu\'imica, Itajub\'a, MG, Brazil}

\author{Amilcar Rabelo de Queiroz}\email{amilcarq@df.ufcg.edu.br}
\affiliation{Universidade Federal de Campina Grande, Campina Grande, PB 58429-900, Brazil}

\author{Elcio Abdalla}\email{eabdalla@usp.br}
\affiliation{Instituto de F\'isica, Universidade de S\~ao Paulo, S\~ao Paulo, Brasil}
\affiliation{Universidade do Estado da Paraíba, R. Baraúnas, 351, Universitário, Campina Grande - PB, 58429-500, Brazil} 
\affiliation{Centro de Ciências Exatas e da Natureza - CCEN\\  Universidade Federal da Paraíba - CEP 58059-970 -- João Pessoa - PB - Brazil }

\date{\today}

\begin{abstract}
The article studies the dynamical behavior of axial gravitational perturbations of homogeneous stars in Anti-de Sitter spacetime. Because the radial coordinate $r$ transforms into the tortoise coordinate $y = y(r)$, obeying $y(r=0)=0$ and $y(r\rightarrow\infty)=y_\text{max}<\infty$, the  tortoise coordinates domain is finite, and the gravitational waves fail to reach out of the domain. Thus, from the perspective of tortoise coordinates, the perturbation equation of uniform density stars in Anti-de Sitter spacetime is analogous to the infinite deep potential well in quantum mechanics. Therefore, the perturbative behavior in this spacetime represents a standing wave vibration. Here we  use shooting method and finite difference method to show the imagery of standing waves in this spacetime for $n=0,1,2$ as examples. On the other hand, by finite difference method, a Gaussian wave packet bounces back and forth within this potential well. Furthermore, due to the shape of the gravitational potential function $V$ within the range $0\le y\le y_\text{max}$, an echo phenomenon occurs at the surface of the star. As the energy of the wave packet remains within the potential well, each time it traverses the surface of the star, a reflective echo is generated. Consequently, after multiple reflections, the Gaussian wave packet cannot maintain its original shape and ultimately disperses. 
\end{abstract}


\maketitle

\section{Introduction}
\renewcommand{\theequation}{1.\arabic{equation}} \setcounter{equation}{0}
Since LIGO announced the detection of gravitational waves generated during the merger of binary black holes \cite{LIGO1}, we officially entered the era of gravitational wave astronomy \cite{LIGO2,LIGO3,LIGO4,LIGO5}. Soon after, gravitational wave detectors also discovered the gravitational waves produced by the merger of binary neutron stars \cite{LIGO6}. As neutron stars can be directly observed through electromagnetic waves, this event marked the first time humans observed the same astronomical phenomenon using two completely different detection media—gravitational waves and electromagnetic waves. Observing this astronomical event signifies that we entered the era of multi-messenger astronomy \cite{LIGO6,LIGO7}.

The study of neutron stars, for example, is of paramount importance, comparable to the study of black holes \cite{Luciano,Werner,Morison,Wheeler,Steven}. Neutron stars represent the ultimate evolutionary fate of medium- to massive-sized stars. During the late stages of their evolution, such stars undergo spectacular supernova explosions, leaving behind a dense core that forms a neutron star \cite{Luciano,Werner,Morison}. Therefore, investigating neutron stars provides critical insights into the detailed processes of stellar evolution. Once the mass of a compact star exceeds the Chandrasekhar limit, it will become a neutron star, and a compact star exceeding the Oppenheimer-Volkoff limit will collapse into a black hole\cite{Steven}. Essentially, The matter composing a neutron star is in a neutron-degenerate state, and it possesses strong gravitational forces\cite{Steven}, which is counterbalanced by the degeneracy pressure of neutrons. Some theorists have even hypothesized that the extreme pressures within a neutron star might compress matter into a quark state\cite{QStar1,QStar2,QStar3}. As a result, neutron stars hold profound significance not only in the field of astronomy but also in particle physics\cite{Neutron}. Similar discussions can be done for other types of stars like white dwarfs, supermassive stars, etc.

The internal structure of stars is profoundly influenced by their equation of state (EOS) \cite{Wheeler,Steven,Neutron,Star1,Star2,SQNMs1,SQNMs2,SQNMs3,SQNMs4}. However, it is challenging to derive the EOS of fundamental particles under extreme conditions of high temperature and pressure directly from first principles. Consequently, empirical formulas established through high-energy physics experiments play a crucial role in studying the detailed internal structure of stars. Nevertheless, if we set aside detailed discussions of the complex internal structures of dense stars, simplified EOS models can provide approximate relationships between the matter density and pressure. Despite such simplifications, obtaining analytical solutions for the internal structure of stars is still challenging. To date, in the study of neutron stars, it is difficult to obtain analytical solutions for neutron star spacetimes in many complex cases. However, in a simple model—uniform density star model—analytical solutions can be obtained\cite{Steven}. Although such a model is over-simplified, its analytical tractability allows for numerous meaningful analyses. Research on uniform-density stars and their dynamical stability provides valuable qualitative insights into the properties of stars. 

Einstein, firmly believing in a static universe, introduced the cosmological constant $\Lambda$ into his theory of general relativity. When $\Lambda > 0$, it provides a repulsive force that Einstein hoped would counteract the inward pull of gravitational attraction\cite{Cosm1}. However, this static model of the universe was definitively disproved when Hubble later discovered that the universe is expanding\cite{Cosm2}. Interestingly, at the end of the 20th century, the discovery of dark energy driving the accelerated expansion of our universe brought the cosmological constant model back into prominence as the most successful dark energy model\cite{Cosm2}. When $\Lambda \neq 0$, the corresponding stationary or static vacuum solutions typically exhibit divergences at infinity. For $\Lambda > 0$, the spacetime described by such solutions is known as the de Sitter spacetime, whereas for $\Lambda < 0$, the spacetime is known as the Anti-de Sitter (AdS) spacetime. 
Generalizations on AdS and dS spacetimes have been addressed more recently. Introducing a fluid in spacetime, the authors have obtained black hole, cosmological universe, and wormhole solutions \cite{Molina,GL1,GL2}. Giambo and Luongo have argued in \cite{GL1} that some of these solutions obtained suffer from instabilities because the sound velocity in the fluid becomes imaginary. 
On the gravitational stability of pure AdS and dS spacetimes, it is relevant to say that in the linear regime both are stable. However, in the non-linear regime AdS spacetime becomes unstable as shown in \cite{Bizon}. On the other hand, Choptuik declares in \cite{Choptuik} that boson stars in AdS can be stable depending on the perturbation. To our knowledge, no analysis of AdS star-like solutions using matching conditions has been carried out.

Furthermore, theorists have discovered a correspondence between the properties of Anti-de Sitter spacetime and conformal field theories, known as the AdS/CFT correspondence \cite{AdSCFT1}. This relationship is believed to hint at a deeper physical significance, potentially offering new perspectives for studying quantum field theory and shedding light on the fundamental nature of gravity. Therefore, the study of perturbations in starlike or non-black hole solutions in the AdS/CFT context can give new insights about the holographic principle and its applications. 

This paper focuses on the axial gravitational perturbations of uniform-density stars in Anti-de Sitter (AdS) spacetime background where the nature of the AdS phase, here, is associated with a constant energy density that may be seen as an extra geometrical contribution to the energy momentum tensor, as suggested by cosmological scenarios, external fields, dark matter contributions, and so on. Although the radial coordinate in this spacetime ranges from $0$ to infinity, the corresponding tortoise coordinate confines the radial domain to a finite range. This restriction implies that the dynamics of gravitational waves in AdS spacetime differs significantly from those in standard spacetimes. In section 2, we derive the uniform-density star solution in AdS spacetime and discuss its properties. We then formulate the perturbation equation for axial gravitational waves. By employing a separation of variables, we transform the perturbation equation into the standard form of the radial Schrödinger equation. In section 3, we investigate axial gravitational waves in AdS spacetime using the shooting method and finite difference method(FDM). Due to the finite effective range of the tortoise coordinate, standing waves arise in this spacetime. We will illustrate the standing wave patterns for $n=0,1,2$. Additionally, the discontinuity at the stellar surface causes a Gaussian wave packet, which reflects repeatedly within the effective range of the tortoise coordinate, to generate regular echo phenomena. These echo patterns will also be presented. Section 4 concludes with a summary and discussion of the results.

\section{main equations of Axial gravitational perturbation in AdS uniform density star}
\renewcommand{\theequation}{2.\arabic{equation}} \setcounter{equation}{0}

As mentioned in the Introduction, the primary advantage of the static uniform-density star model lies in the analytical solution. This solution enables a convenient and qualitative discussion of the key characteristics of the spacetime surrounding stars. It is well-known that the metric of a static, spherically symmetric spacetime can generally be expressed as \cite{Steven}
 \bqn
 \lb{Metric1}
ds^2&=&-f(r)dt^2+\frac{dr^2}{h(r)}+r^2d\theta^2+r^2\sin^2\theta d\varphi^2.~~~
 \eqn
Outside the star, where $r>r_s$ ($r_s$ being the radius of the stellar surface), the energy-momentum tensor vanishes there.

Inside the star, where $r<r_s$, the matter field can be treated as an ideal fluid. The corresponding energy-momentum tensor can be written as
 \bqn
 \lb{Metric2}
T^m_{\mu\nu}=\left[\rho+P(r)\right]u_\mu u_\nu+P(r) g_{\mu\nu}\quad .
 \eqn
Here, $\rho$ represents the density of the stellar matter, which is a constant in the uniform-density star model. $P(r)$ denotes the pressure of the fluid, and the 4-velocity is
 \bqn
 \lb{Metric3}
u_\mu=\frac{1}{\sqrt{f(r)}}\delta_\mu^t,
 \eqn
so that $g^{\mu\nu}u_\mu u_\nu=-1$.

To calculate the metric solution for the spacetime of a uniform-density star, we substitute the above relations into ``geometrically modified" Einstein's field equations by a cosmological constant term,
 \bqn
 \lb{fieldEqu1}
R_{\mu\nu}-\frac{1}{2}g_{\mu\nu}R&=&8\pi G_n T_{\mu\nu}-g_{\mu\nu}\Lambda.
 \eqn
Clearly, outside of the star, the static solution to Einstein's field equations is the Schwarzschild-AdS solution, and its metric is given by:
 \bqn
 \lb{BackGroundSolution1}
 \text{as}&~&r>r_s\nb\\
f_o(r)&=&h_o(r)=1-\frac{2G_nM}{r}-\frac{\Lambda}{3}r^2,\nb\\
\rho_o&=&P_o(r)=0.
 \eqn
On the other hand, inside the uniform density star, the solution is given by
 \bqn
 \lb{BackGroundSolution2}
  \text{as}&~&r<r_s\nb\\
f_i(r)&=&\left[\frac{\bar{\Lambda}}{\hat{\Lambda}}\sqrt{1-\frac{\hat{\Lambda}}{3}r^2}+\frac{12G_n\pi\rho}{\hat{\Lambda}}\sqrt{1-\frac{\hat{\Lambda}}{3}r_s^2}\right]^2,\nb\\
h_i(r)&=&1-\frac{1}{3}\hat{\Lambda}r^2,\nb\\
P_i(r)&=&\frac{\sqrt{1-\frac{\hat{\Lambda}}{3}r_s^2}-\sqrt{1-\frac{\hat{\Lambda}}{3}r^2}}{12G\pi\rho\bar{\Lambda}^{-1}\sqrt{1-\frac{\hat{\Lambda}}{3}r_s^2}+\sqrt{1-\frac{\hat{\Lambda}}{3}r^2}}\rho,
 \eqn
where $\bar{\Lambda}\equiv\Lambda-4G_n\pi\rho$, $\hat{\Lambda}\equiv\Lambda+8G_n\pi\rho$ and $M\equiv\frac{4}{3}\pi\rho r_s^3$.

From the solution of $h_i(r)$, it can be observed that when $\Lambda>0$, the effect of cosmological constant plays the role of a repulsive force. This is the reason Einstein introduced this constant to provide a repulsive force in his cosmological model. The effect of positive cosmological constant can (partially) substitute for the positive pressure $P_i$ inside the star. In contrast, when $\Lambda<0$, the cosmological constant generates an inward gravitational force, which leads to the need for a larger $P_i$ to counteract the effect. The Fig. \ref{FigP} shows the relation between $P_i(r)$ and $\Lambda$.

\begin{figure}[htbp]
\centering
\includegraphics[width=0.7\columnwidth]{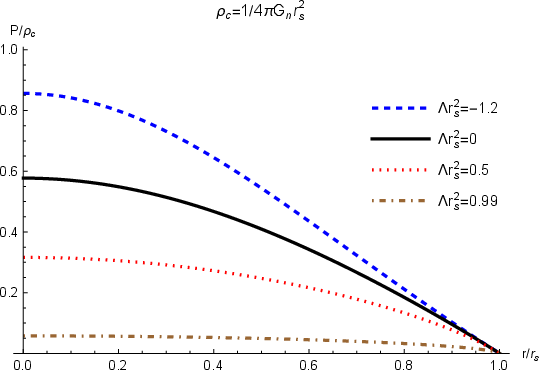}
\caption{the relation between $P_i(r)$ and $\Lambda$}
\lb{FigP}
\end{figure}

At the boundary, where $r=r_s$, the junction conditions are satisfied:
 \bqn
 \lb{BackGroundSolution3}
f_i(r_s)&=&h_i(r_s)=f_o(r_s)=h_o(r_s),\nb\\
P(r_s)&=&0.
 \eqn
Therefore, for the entire spacetime, $f$, $h$, $P$ and $\rho$ can be written as:
 \bqn
 \lb{BackGroundSolution4}
f(r)&=&[1-U(r-r_s)]f_i(r)+U(r-r_s)f_o(r),\nb\\
h(r)&=&[1-U(r-r_s)]h_i(r)+U(r-r_s)h_o(r),\nb\\
P(r)&=&[1-U(r-r_s)]P_i(r),\nb\\
\rho(r)&=&[1-U(r-r_s)]\rho_i,
 \eqn
where
 \bqn
 \lb{BackGroundSolution4}
U(r-r_s)=\left\{\begin{array}{ccc}
        0 & & r<r_s \\
        1 & & r>r_s
        \end{array}\right..
 \eqn

In this paper, without loss of generality, we choose $8\pi G_n=1$, $r_s=1$, which is equivalent to the following dimensionless transformation:
 \bqn
 \lb{dimensionless1}
r&\to& \tilde{r}=r/r_s,\nb\\
t&\to& \tilde{t}=t/r_s,\nb\\
\omega&\to& \tilde{\omega}=\omega\times r_s,\nb\\
\rho&\to& \tilde{\rho}=\rho\times r_s^2,\nb\\
\Lambda&\to& \tilde{\Lambda}=\Lambda\times r_s^2
 \eqn
the transformation can simplify our research in the paper. It should be noted that, due to this transformation, $\Lambda$ is no longer considered a fixed constant. In the following calculations, we drop the tilde symbols on the above variables.

In Fig. \ref{FigMetric}, we show the graphs of $f(r)$, $h(r)$, and $P(r)$ for three different values of density, $\rho = 0.5$, $1.5$, and $2.5$. We observe that as the density $\rho$ increases, the pressure at the center of the star ($r = 0$) also increases. This is because, with a constant volume, as $\rho$ increases, the mass of the star increases, and to maintain the balance of forces inside the star, the internal pressure of the star must increase. Furthermore, the difference between $f(0)$ and $h(0)$ at the center becomes more pronounced.
\begin{figure}[htbp]
\centering
\includegraphics[width=0.33\columnwidth]{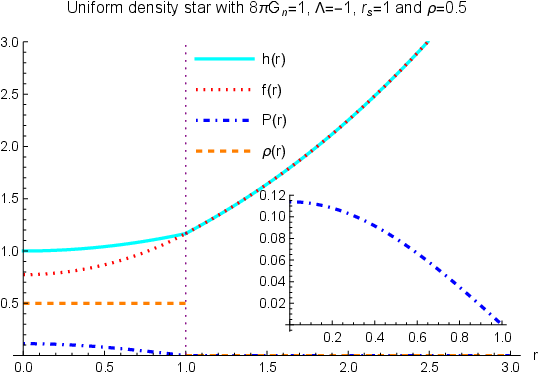}\includegraphics[width=0.33\columnwidth]{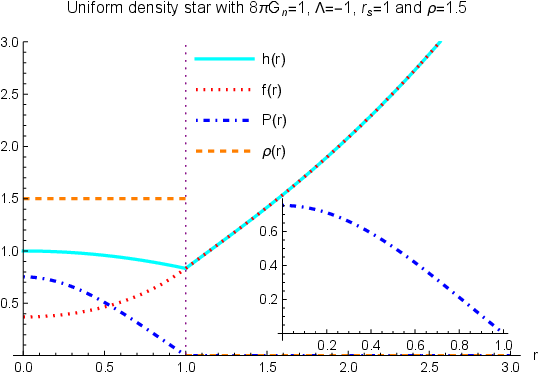}\includegraphics[width=0.33\columnwidth]{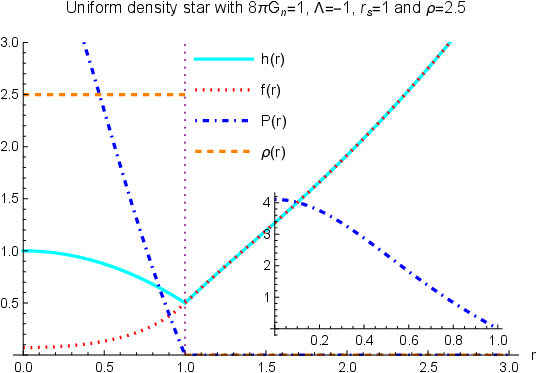}
\caption{The metric of uniform density stars in Anti-de Sitter spacetime, where $P(r)$ and $\rho(r)$ are the pressure and the density of the star respectively, and the vertical dashed line indicates the position of the star's surface.}
\lb{FigMetric}
\end{figure}

In this static background spacetime, a gravitational perturbation is introduced, which corresponds to a gravitational wave. Gravitational waves have two polarization directions. In order to obtain decoupled gravitational perturbation equations in a static spherically symmetric spacetime, Regge and Wheeler proposed the famous Regge-Wheeler gauge \cite{Regge}. In this gauge, gravitational perturbations can be divided into axial and polar perturbations. This paper studies the axial gravitational perturbation, for which the following ansatz is chosen:
\bqn
\lb{odd2}
\delta g_{\mu\nu}&=&\sin\theta\partial_\theta P(\cos\theta)\left(\begin{array}{cccc}
        0 & 0 & 0 & h_0(r) \\
        0 & 0 & 0 & h_1(r) \\
        0 & 0 & 0 & 0 \\
        h_0(r) & h_1(r) & 0 & 0
        \end{array}\right)e^{-i\omega t},\nb\\
\delta u_\mu&=&-e^{-i\omega t}\delta_\mu^\varphi h_w(r)\sin^{-2}\theta\partial_\theta P(\cos\theta),\nb\\
\delta p&=&-e^{-i\omega t}\delta_\mu^\varphi p_w(r)\sin\theta\partial_\theta P(\cos\theta),\nb\\
\delta \rho&=&-e^{-i\omega t}\delta_\mu^\varphi \rho_w(r)\sin\theta\partial_\theta P(\cos\theta).
\eqn
where $P(\cos\theta)$ is the Legendre functions.

Submitting above ansatz into gravitational field equation and the conservation of the stress–energy tensor, we find the relation of $h_w$, $h_1$ and $h_0$ as follows
 \bqn
 \lb{odd2a}
h_w(r)&=&p_w(r)=\rho_w(r)=0,\nb\\
h_0(r)&=&\frac{f}{i\omega r}\left\{\left[f-1+\Lambda r^2-4G\pi r^2\left(3p+\rho\right)\right]h_1-rhh_1'\right\}.
 \eqn
The results imply that the perturbation of fluid won't affect the results of axial perturbation, and we only need to study $h_1(r)$ to understand this perturbation. In order to simplify equation, we can set
 \bqn
 \lb{odd3}
h_1(r)=B(r)\Phi(r),
 \eqn
where $B(r)$ satisfies the equation
 \bqn
 \lb{odd4}
rhB'+\left[1-\Lambda r^2-2h+4G\pi (P-\rho)r^2\right]B=0.
 \eqn
Substituting the ansatz (\ref{odd2}) and (\ref{odd3}) into gravitational field equation (\ref{fieldEqu1}), we can obtain the perturbation equation
 \bqn
 \lb{perturbation1}
\frac{d^2\Phi}{dy^2}+\left[\omega^2-V(r)\right]\Phi=0.
 \eqn
In the perturbation equation, $y\equiv\int_0^r\frac{dr}{\sqrt{f(r)h(r)}}$ is the tortoise coordinate, and the potential function is given by
 \bqn
 \lb{perturbation2}
V=f\left[\frac{L^2+L-3+3h}{r^2}+\Lambda-4\pi G(P-\rho)\right],~~~
 \eqn
where $L$ is the angular quantum number originating from Legendre functions $P(\cos\theta)$.

\begin{figure}[htbp]
\centering
\includegraphics[width=0.35\columnwidth]{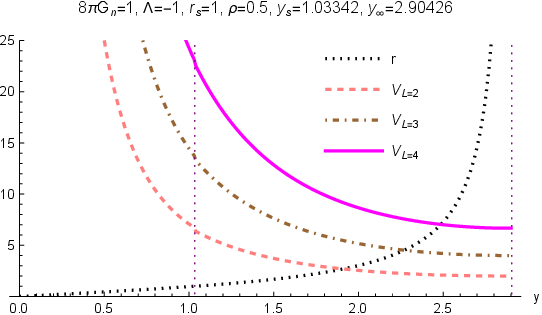}\includegraphics[width=0.35\columnwidth]{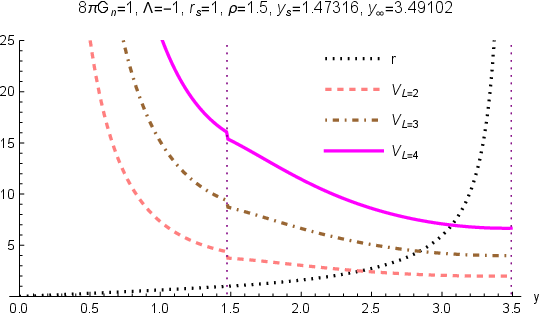}\includegraphics[width=0.35\columnwidth]{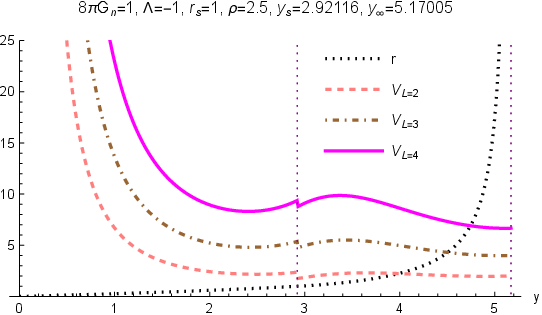}
\caption{Axial gravitational potentials and tortoise coordinates of uniform density star in Anti-de Sitter spacetime, where the vertical dashed line indicates the position of the star's surface.}
\lb{FigPotential}
\end{figure}
In Fig. \ref{FigPotential}, we plot the graphs of $V=V(y)$ and $r=r(y)$ with the tortoise coordinate $y$ as the horizontal axis. From this figure, it can be seen that $y(0)=0$ and $y(r\rightarrow\infty)=y_\text{max}<\infty$, thus the range of the tortoise coordinate is finite. This characteristic leads to a special form of gravitational waves in this spacetime, which we will study in the next section.

\section{Numerical Computation Using the Shooting Method and Finite Difference Method}
\renewcommand{\theequation}{3.\arabic{equation}} \setcounter{equation}{0}

From the form of the perturbation equation (\ref{perturbation1}), it can be seen as a standard one-dimensional stationary Schrödinger equation. Mathematically, this is an eigenvalue equation in the frequency domain. By performing a Fourier inverse transform and using the mathematical relation $-i\omega\leftrightarrow\partial_t$, this equation can be converted in the time domain, in the form
 \bqn
 \lb{perturbation3}
-\frac{\partial^2\Psi}{\partial t^2}+\frac{\partial^2\Psi}{\partial y^2}-V(r)\Psi=0\quad .
 \eqn
This is a wave equation with potential for a field that propagates at the speed of light $c=\frac{dy}{dt}=1$. Since the range of the tortoise coordinate is $0\le y\le y_\text{max}$, this means that within a finite time the perturbation can reach the point $y_\text{max}$, which corresponds to the asymptotic region of the Anti-de Sitter spacetime as $r\rightarrow\infty$. Therefore, when $r$ is sufficiently large, the speed of gravitational waves in this spacetime diverges, $v\equiv\frac{dr}{dt}\rightarrow\infty$. 
It should be clarified that $v$ represents a coordinate velocity, and its divergence stems from the asymptotic behavior of the AdS spacetime metric as 
$r\to\infty$ rather than indicating the actual physical propagation speed of gravitational waves.
As a result, within a finite time, a wave packet can appear at any point between $0\le y\le y_\text{max}$, which also implies that $\Psi(y<0)=\Psi(y>y_\text{max})=0$, meaning that the boundary conditions require
 \bqn
 \lb{perturbation4}
\Psi(y=0)&=&\Psi(y_\text{max})=0.
 \eqn
This is exactly the Schrödinger equation for a one-dimensional infinite potential well, along with its boundary conditions. However, due to the complexity of the potential function, it is hard to obtain an analytical solution for this equation. We use numerical methods to solve it.

For the frequency-domain equation (\ref{perturbation1}), an effective method to numerically solve, for the eigenvalues and eigenfunctions, is the Shooting method. Since $y=0$ is a regular singular point, we can expand $\Psi$ in a series around it and calculate the series expansion solution for $\Psi$ near $y=0$, which includes an undetermined parameter $\omega$ \cite{SQNMs1,SQNMs2,SQNMs3,SQNMs4}. Then, we use this series solution $\Psi(\delta)$ and $\Psi'(\delta)$ (where $0 < \delta \ll 1$) as boundary values for the numerical computation and attempt to adjust $\omega \in \mathbb{R}$ such that $\Psi(y_\text{max}) = 0$. The value of $\omega$ that satisfies this condition is the eigenvalue, and the corresponding $\Psi$ is the eigenfunction. During the solution process, it is important to note that at the surface of the star, $y = y_s$, the matching conditions must be satisfied,
 \bqn
 \lb{perturbation5}
\Psi_i(y_s)&=&\Psi_o(y_s),\nb\\
\partial_y\Psi_i(y_s)&=&\partial_y\Psi_o(y_s).
 \eqn
During the computation of normal mode frequencies via the shooting method, we adopted several verification techniques outlined in \cite{MethodLK} to ensure the validity and reliability of our results.

\begin{figure}[t]
\centering
\includegraphics[width=1\columnwidth]{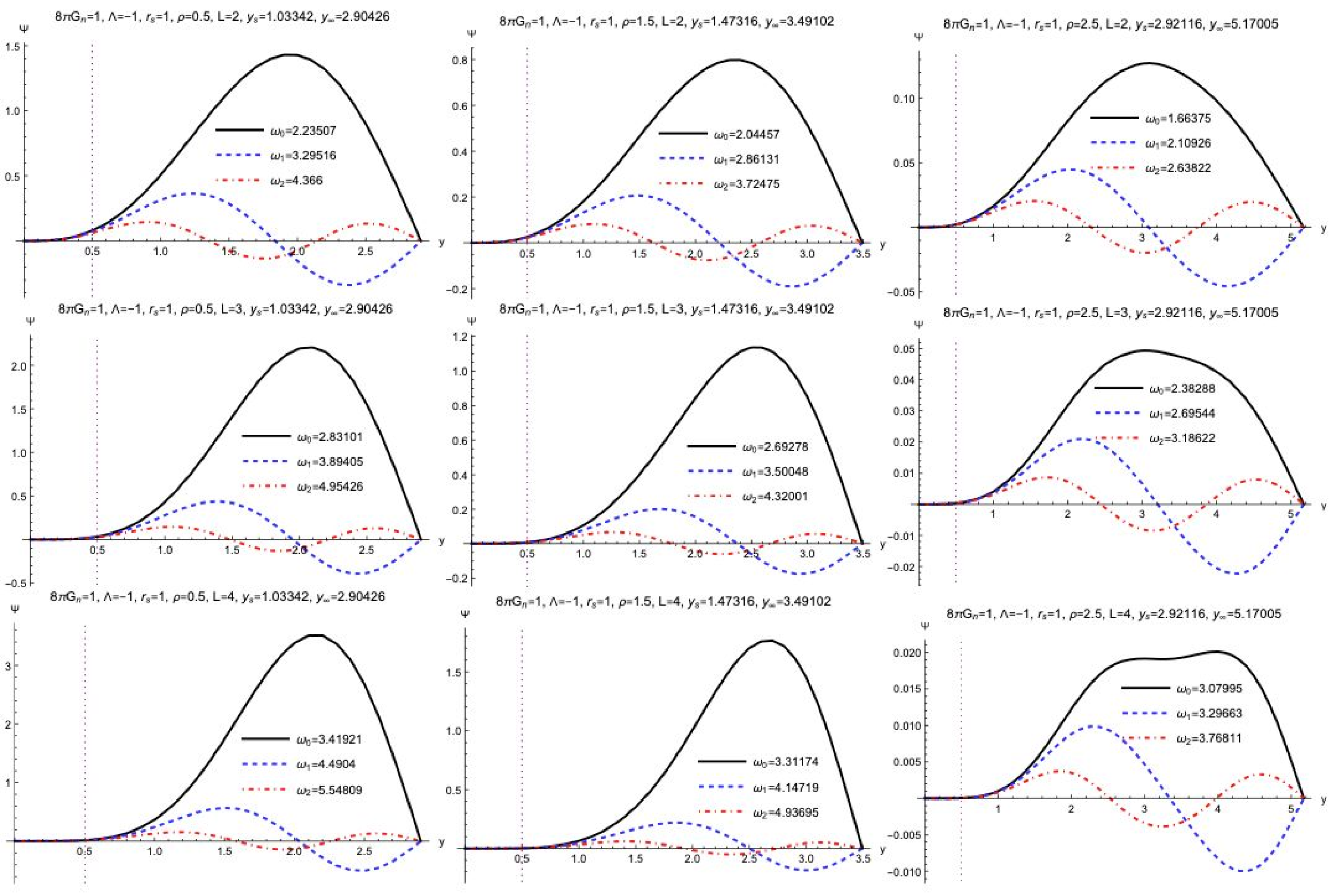}
\caption{Axial gravitational normal modes with their respective eigenfunctions and eigenvalues, where the vertical dashed line indicates the position of the star's surface.}
\lb{FigNM}
\end{figure}

In Fig. \ref{FigNM}, we present the eigenvalues and eigenfunctions obtained using the shooting method for $\Lambda=-1$, $\rho=0.5,1.5,2.5$ and $n=0,1,2$. It is evident that in the present case, $\omega$ is a real number. This is because $y\in[0,y_\text{max}]$, so in the tortoise coordinates, the energy carried by the wave packet can only bounce back and forth within this range and will not propagate to $y\rightarrow\pm\infty$. Therefore, the energy does not dissipate, which means $Im(\omega)=0$. This is different from the case of gravitational waves in ordinary black hole spacetimes and star spacetimes. Thus, the disturbances treated in this paper are not quasinormal mode oscillations but rather normal mode oscillations \cite{NormalModes1, NormalModes2}.

\begin{figure}[t]
\centering
\includegraphics[width=1\columnwidth]{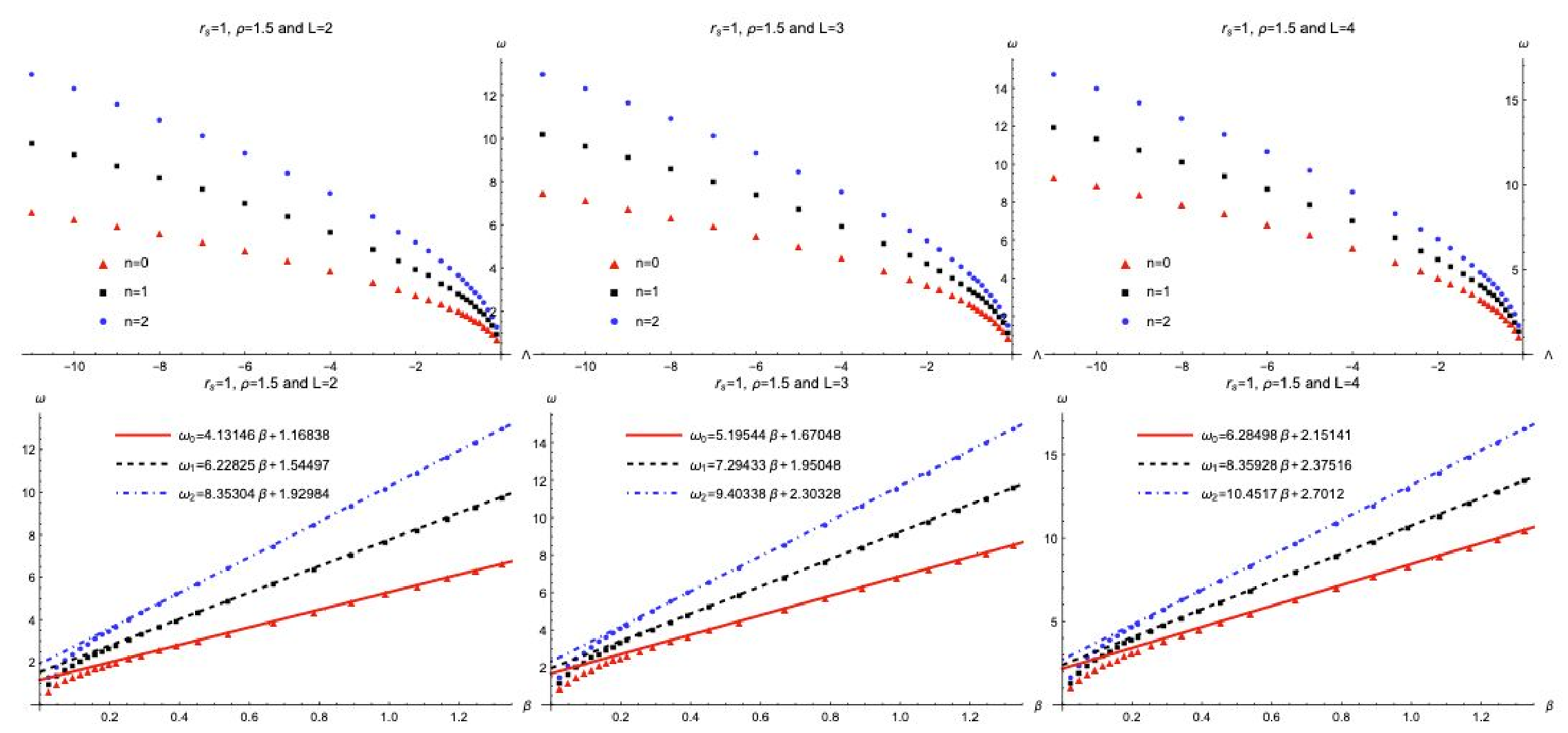}
\caption{The relationship between the  axial gravitational normal mode frequencies $\omega$, the cosmological constant $\Lambda$, and the Schwarzschild radius $r_\text{Sch}$, where $\beta=-\frac{r_\text{Sch}}{r_s}\Lambda r_\text{Sch}$, and Schwarzschild radius $r_\text{Sch}=2G_nM$ satisfies $f_o(r_\text{Sch})=0$, and the vertical dashed line indicates the position of the star's surface.}
\lb{FigWLAMBDA}
\end{figure}

We know that the quasi-normal modes of black holes in Anti-de Sitter spacetime have a very famous property: for large-scale black holes, there exists a linear relationship in the quasi-normal mode frequencies\cite{AdS1,AdS2}. From FIG \ref{FigWLAMBDA}, it shows the fact that a similar linear relationship still exists in the spacetime of compact stars in Anti-de Sitter spacetime.

On the other hand, the finite difference method can also be used to solve the time-domain equation \cite{FDM1, FDM2, FDM3, FDM4, FDM5, FDM6, FDM7}. To do this, we need to discretize equation (\ref{perturbation3}) as
 \bqn
 \lb{perturbation6}
\Psi_j^{k+1}=-\Psi^{k-1}_j+\frac{\Delta t^2}{\Delta y^2}\left(\Psi^k_{j+1}+\Psi^k_{j-1}\right)+\left(2-2\frac{\Delta t^2}{\Delta y^2}+\Delta t^2V_j\right)\Psi^k_j\quad ,
 \eqn
where $\Psi^k_j\equiv\Psi(t=t_k,y=y_j)$ and $V_j\equiv V(y_j)$. $\Delta y$ and $\Delta t$ are the step sizes in the $y-$ and $t-$directions respectively, and satisfy the physical condition
 \bqn
 \lb{perturbation7}
\Psi^k_0&=&\Psi^k_N=0,\nb\\
\Psi^k_s&=&\left(1+\frac{\Delta y_o}{\Delta y_i}\right)^{-1}\left(\Psi^k_{s+1}+\frac{\Delta y_o}{\Delta y_i}\Psi^k_{s-1}\right)\, ,
 \eqn
where $\Psi^k_0\equiv\Psi(t=t_k,y=0)$, $\Psi^k_N\equiv\Psi(t=t_k,y=y_\text{max})$, $\Psi^k_s\equiv\Psi(t=t_k,y=y_s)$, $\Psi^k_{s\pm1}\equiv\Psi(t=t_k,y=y_{s}\pm\Delta y)$, while $\Delta y_o$ and $\Delta y_i$ are the step sizes in $y$ coordinate in the external and internal regions of the star, respectively.

If we choose the results of Fig. \ref{FigNM} as initial condition, we can obtain Fig. \ref{FigStandingL2}, Fig. \ref{FigStandingL3} and Fig. \ref{FigStandingL4}.

\begin{figure}[htbp]
\centering
\includegraphics[width=1\columnwidth]{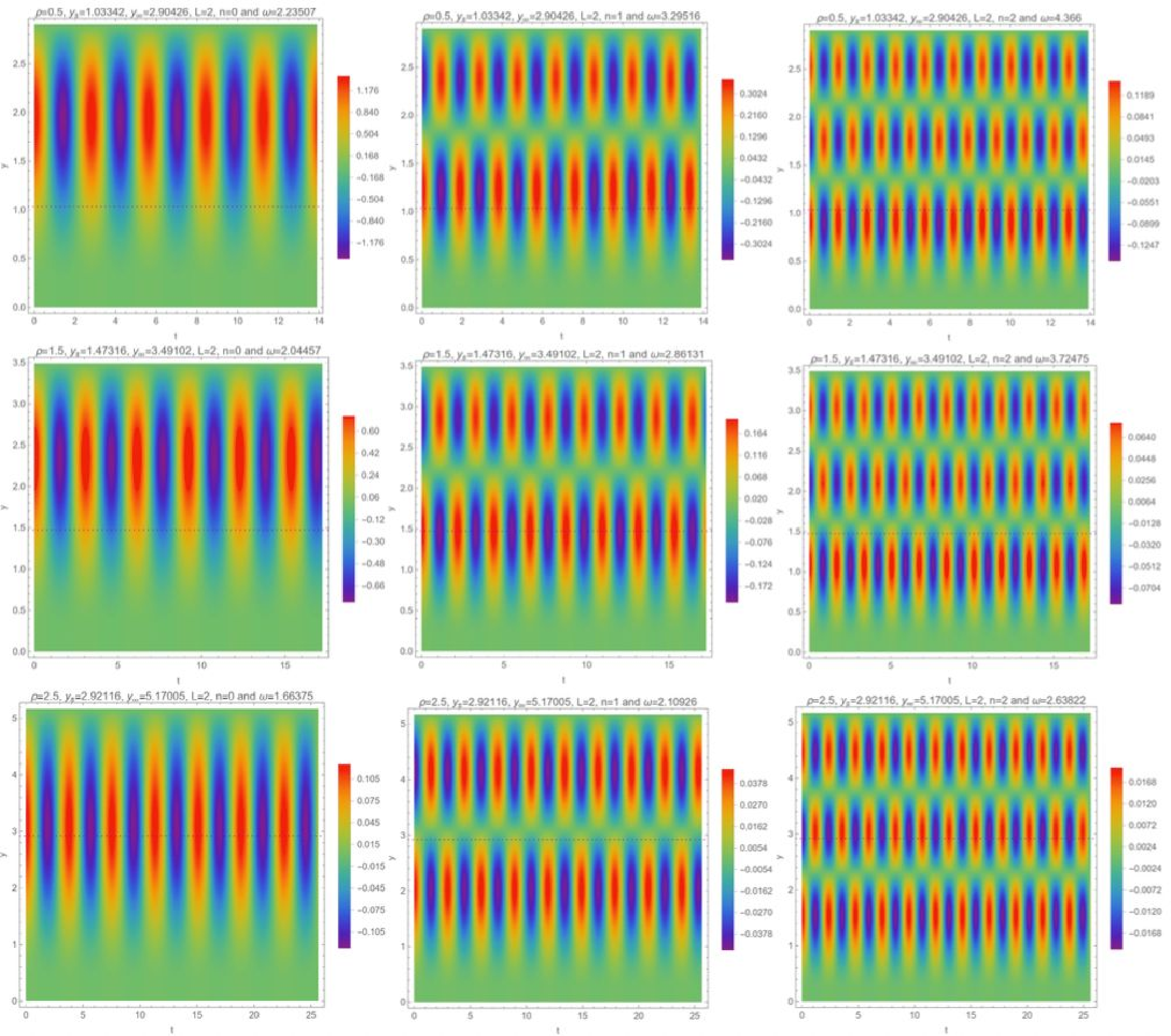}
\caption{Axial gravitational standing wave with $L=2$, where the horizontal dashed line indicates the position of the star's surface.}
\lb{FigStandingL2}
\end{figure}

\begin{figure}[htbp]
\centering
\includegraphics[width=1\columnwidth]{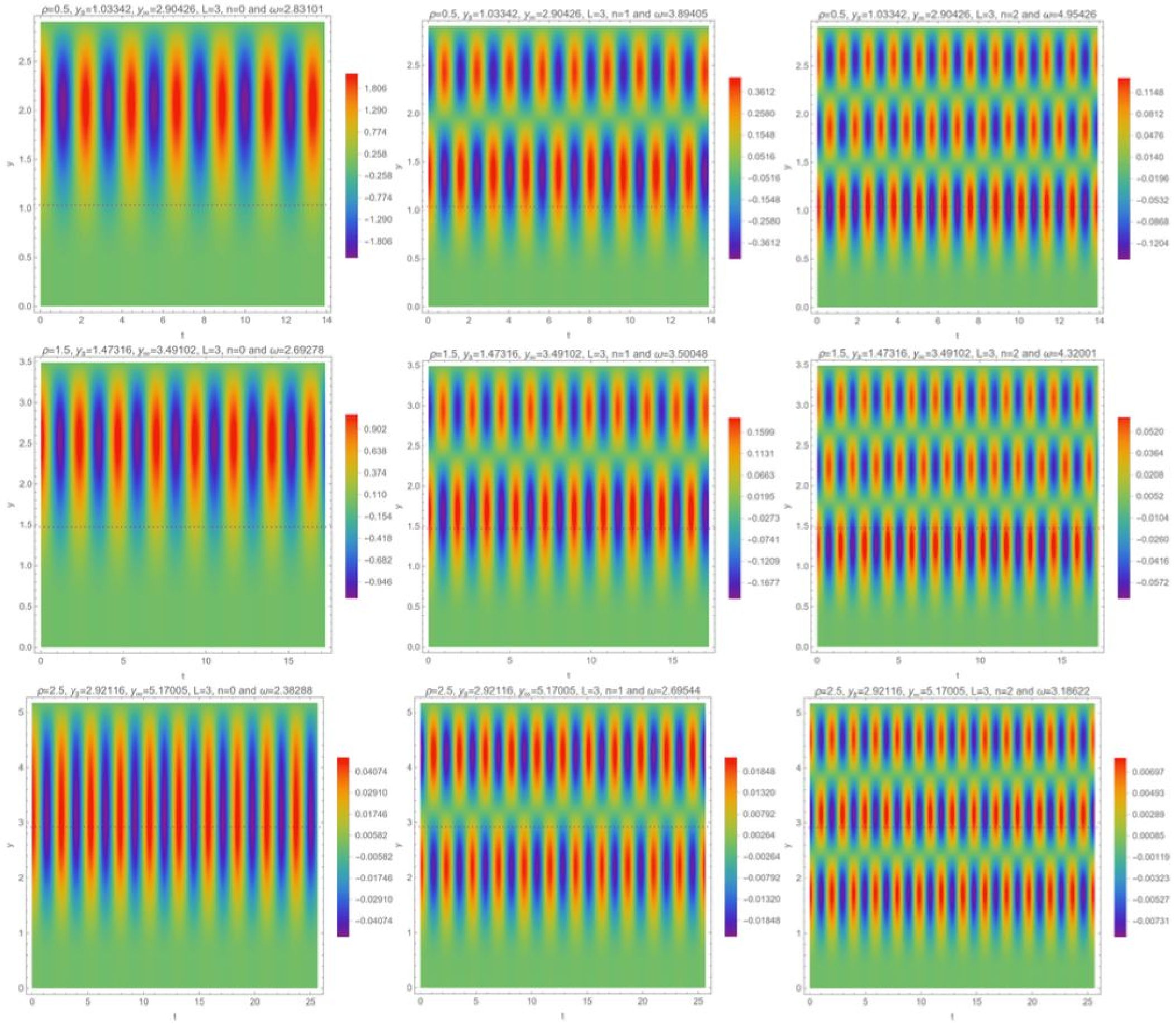}
\caption{Axial gravitational standing wave with $L=3$, where the horizontal dashed line indicates the position of the star's surface.}
\lb{FigStandingL3}
\end{figure}

\begin{figure}[htbp]
\centering
\includegraphics[width=1\columnwidth]{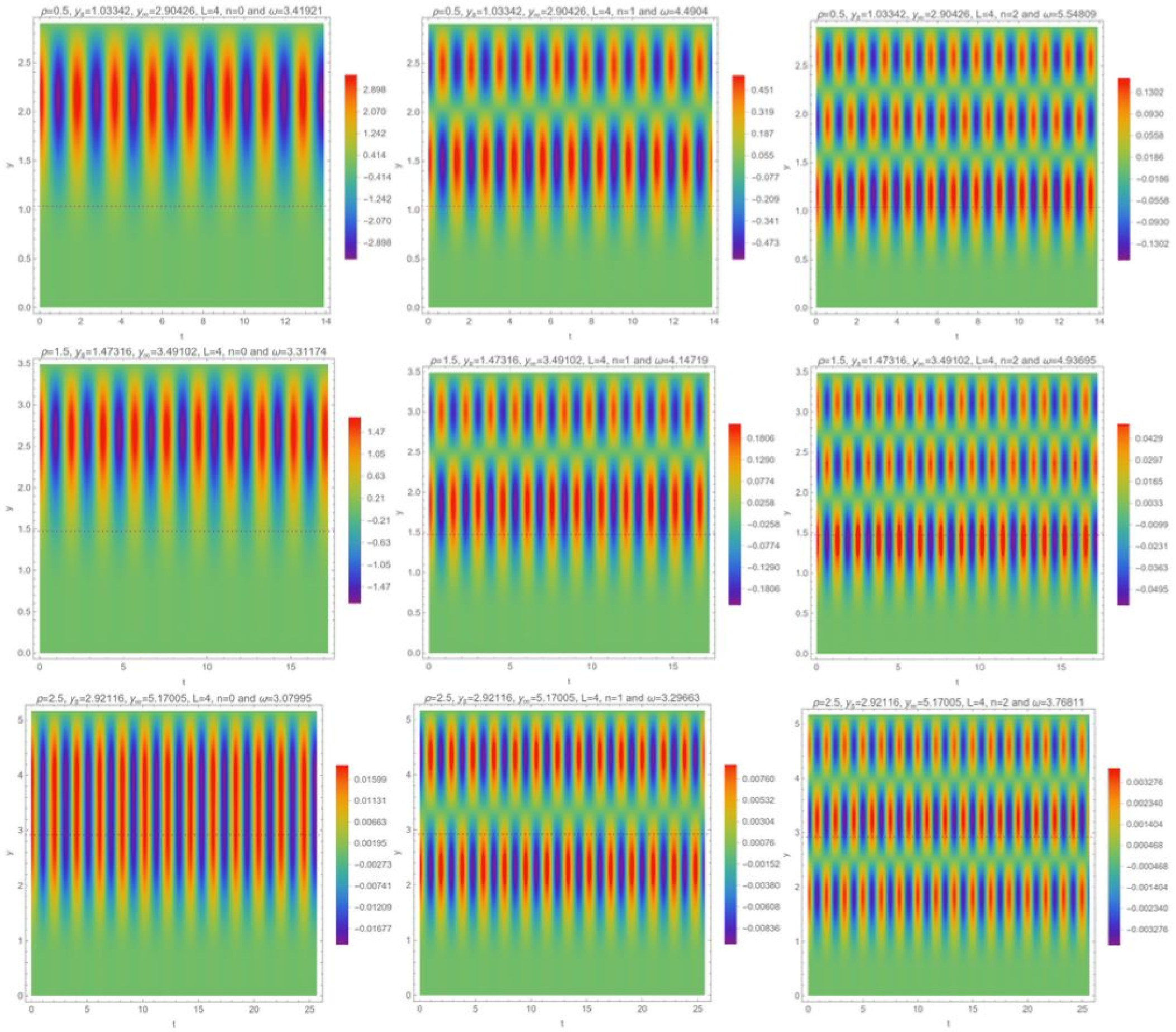}
\caption{Axial gravitational standing wave with $L=4$, where the horizontal dashed line indicates the position of the star's surface.}
\lb{FigStandingL4}
\end{figure}

It is evident that during the dynamics of the wave described above, the positions of the wave crests and nodes do not change with time, and the amplitude of the wave at each point remains constant. This is exactly the characteristic of standing waves. This is because condition (\ref{perturbation4}) ensures that the energy of wave cannot propagate beyond the regions $y<0$ and $y>y_\text{max}$, which are physically inaccessible.

On the other hand, if our initial condition is a Gaussian wave packet $\Psi_0=e^{-\frac{(y-y_o)^2}{(2\sigma)^2}}$, the images are shown in Fig. \ref{FigEchoA}, Fig. \ref{FigEchoB}, and Fig. \ref{FigEchoC}:
\begin{figure}[htbp]
\centering
\includegraphics[width=1\columnwidth]{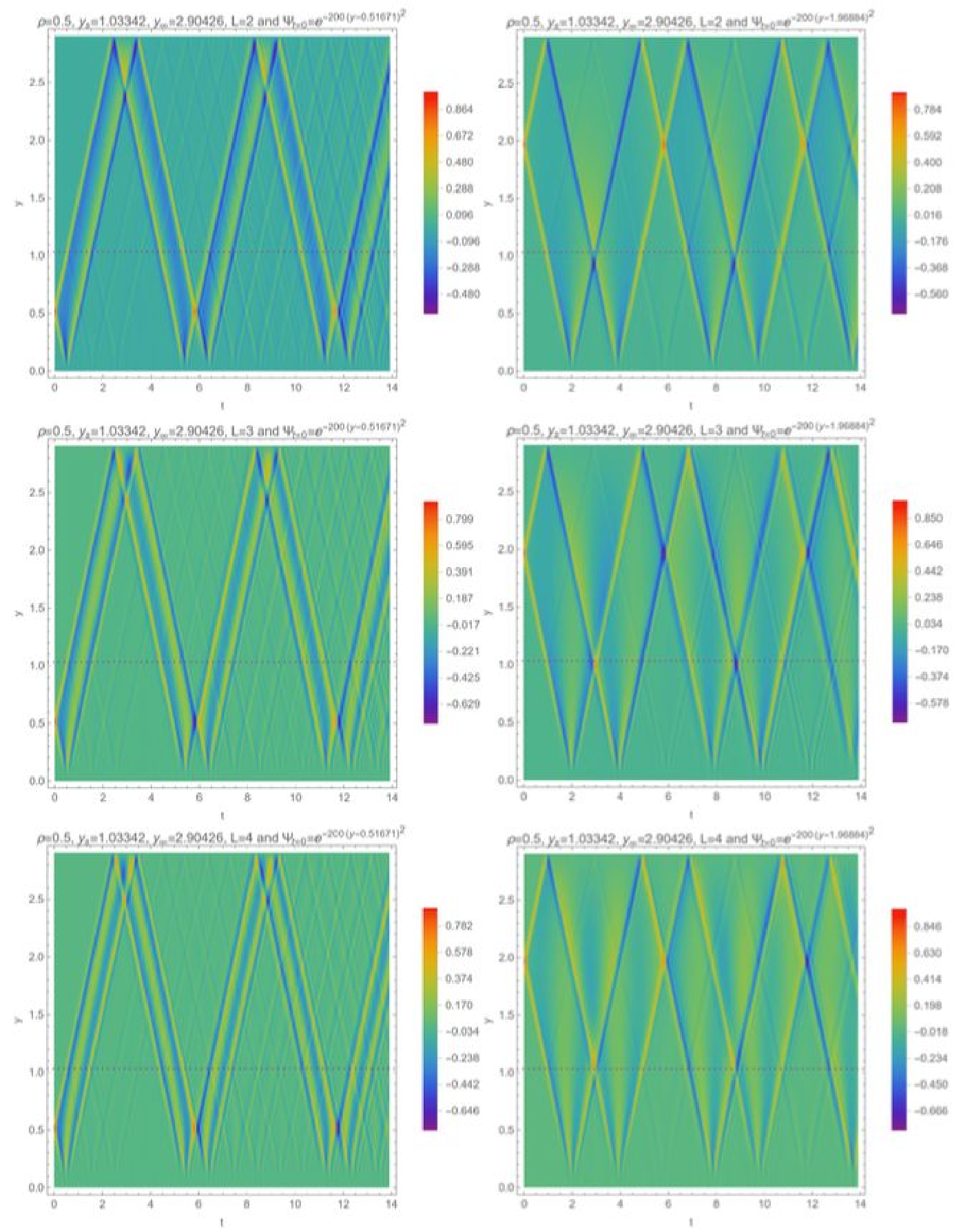}
\caption{Echoes from surface of star with $\rho=0.5$, where the horizontal dashed line indicates the position of the star's surface.}
\lb{FigEchoA}
\end{figure}
\begin{figure}[htbp]
\centering
\includegraphics[width=1\columnwidth]{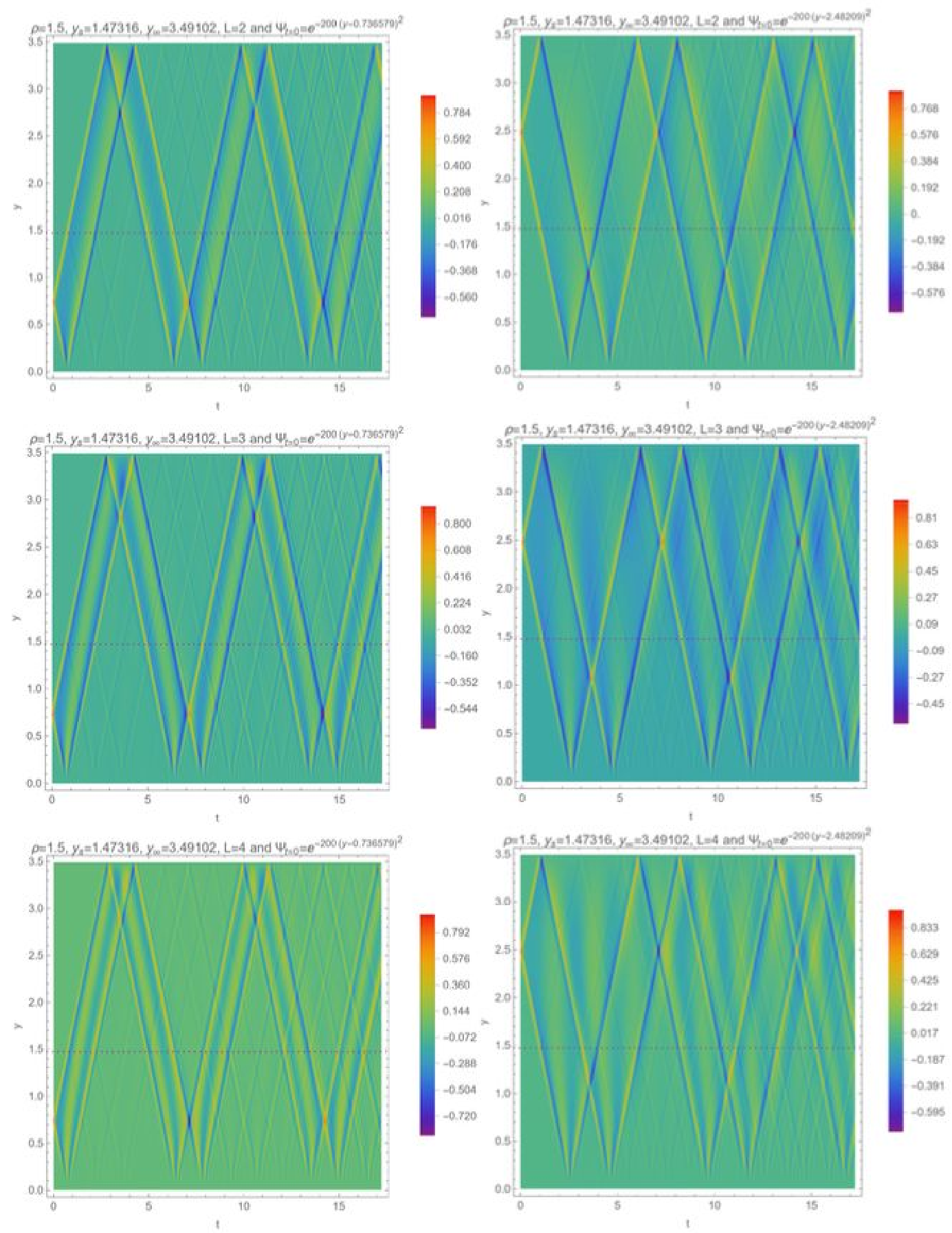}
\caption{Echoes from surface of star with $\rho=1.5$, where the horizontal dashed line indicates the position of the star's surface.}
\lb{FigEchoB}
\end{figure}
\begin{figure}[htbp]
\centering
\includegraphics[width=1\columnwidth]{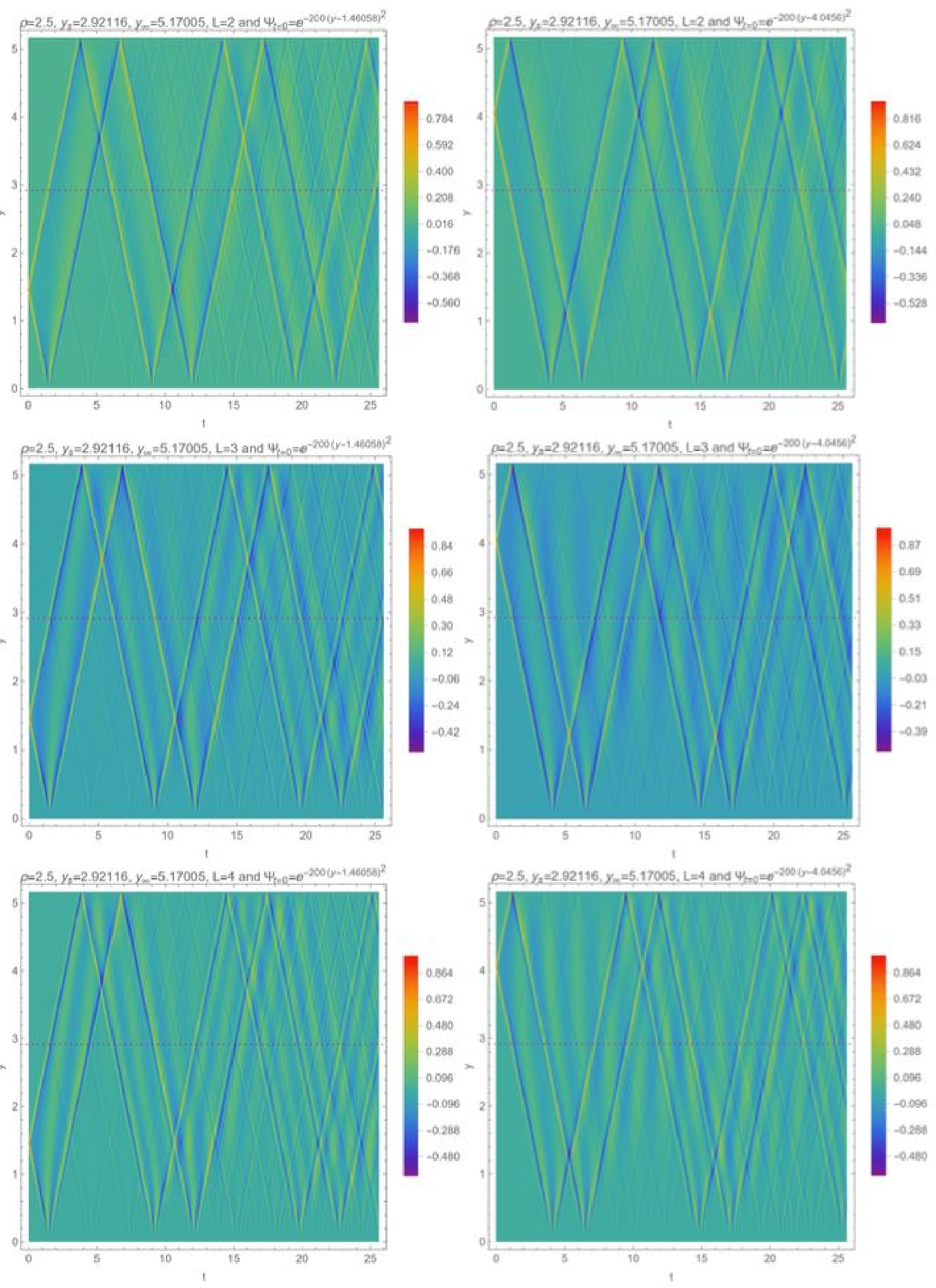}
\caption{Echoes from surface of star with $\rho=2.5$, where the horizontal dashed line indicates the position of the star's surface.}
\lb{FigEchoC}
\end{figure}

As mentioned above, the Gaussian wave packet bounces back and forth between $y=0$ and $y=y_\text{max}$. However, due to the existence of the discontinuity on both sides of the star's surface, both reflected and transmitted waves coexist, leading to the echo phenomenon \cite{echoTwo}. Additionally, since the wave packet in this spacetime bounces back and forth between the two ends, each time it passes the star's surface, new reflected waves are generated. As a result, the energy of the main wave packet gradually decreases and more and more wavelets and sub-wavelets emerge, leading to a decreasing of the amplitude and
an increasing of the number of modes of oscillation.
It looks like a mechanism for decomposition of information into modes of oscillation. 
\section{Conclusion}
\renewcommand{\theequation}{4.\arabic{equation}} \setcounter{equation}{0}
This paper investigates the axial gravitational perturbations of an uniform density star in Anti-de Sitter spacetime. The most notable feature of this spacetime is exhibiting gravitational standing wave vibrations. 
It is worth to mentioning that this result is more related to the nature of spatial infinity in asymptotically AdS spacetime than the coordinate used to represents the phenomenon. However it become more clear and illustrative when expressed in tortoise coordinate, which makes this problem analogous to the infinite potential well problem in quantum mechanics.
On the other hand, due to the discontinuity at the surface of the star, gravitational waves in the form of a Gaussian wave packet reflect off the star's surface, carrying away some energy and generating echo phenomena. Furthermore, since the wave packet can only bounce back and forth between $y=0$ and $y=y_\text{max}$, each bounce generates more reflected waves, causing the energy of the main wave packet to gradually disperse across the smaller waves.
Due to the AdS-CFT correspondence, the properties of AdS spacetime are theoretically very important. This paper demonstrates that the properties of relativistic compact stars in AdS spacetime are relatively simple, which may be helpful for research on AdS spacetime. Since the study of this spacetime is confined to a finite region in the tortoise coordinate, compared to AdS black hole spacetimes, the numerical calculations require significantly fewer resources. These advantages make the study of this spacetime sufficiently meaningful.

\appendix

\section{Scalar perturbation of pure Anti-de Sitter Spacetime}
\renewcommand{\theequation}{A.\arabic{equation}} \setcounter{equation}{0}
The simplest perturbation of a uniform density star is difficult to solve analytically. However, for the pure Anti-de Sitter spacetime, the properties of its perturbation equations are very similar to the situation in this paper, making it highly valuable for reference.

The metric of pure Anti-de Sitter spacetime is given by
 \bqn
 \lb{AdSSpacetime1}
ds^2&=&-F(r)dt^2+\frac{dr^2}{F(r)}+r^2d\theta^2+r^2\sin^2\theta d\varphi^2,\nb\\
F(r)&=&1-\frac{\Lambda}{3}r^2.
 \eqn
Without loss of generality, we set $\Lambda=-3$, so the radial scalar perturbation equation in this spacetime is:
 \bqn
 \lb{AdSSpacetime2}
&&\frac{\partial^2\Phi}{\partial y^2}-\frac{\partial^2\Phi}{\partial t^2}-\left[2\sec^2(y)+L(L+1)\csc^2(y)\right]\Phi=0,
 \eqn
and the tortoise coordinate is given by
 \bqn
 \lb{AdSSpacetime3}
y=\arctan(r).
 \eqn
Here, the tortoise coordinate $y\in[0,\pi/2]$, with $y(r=0)=0$ and $y(r\rightarrow\infty)=\pi/2$. Therefore, the domain of the equation in the tortoise coordinate is confined to a finite range and satisfies:
 \bqn
 \lb{AdSSpacetime4}
\Phi(y=0)=\Phi(y=\pi/2)=0.
 \eqn
 
Therefore, the analytical solution to equation (\ref{AdSSpacetime2}) can be obtained as:
 \bqn
 \lb{AdSSpacetime5}
\Phi&=&\sum_nC_n\Phi_n,\nb\\
\Phi_n&=&-\Phi_Ce^{-i(2n+L+3)t}\cos^2(y)\left[-\sin^2(y)\right]^{\frac{L+1}{2}}{}_2F_1\left[-n,L+3+n,\frac{5}{2},\cos^2(y)\right]\nb\\
&=&-\Phi_Ce^{-i(2n+L+3)t}\left(-\frac{r^2}{1+r^2}\right)^{\frac{L+1}{2}}\frac{1}{\left(1+r^2\right)}{}_2F_1\left[-n,L+3+n,\frac{5}{2},\frac{1}{1+r^2}\right].
 \eqn
Here, $n$ and $L$ are the principal quantum number and the angular quantum number, respectively. It is easy to notice that as $r\rightarrow\infty$, the wave speed $v=\frac{dr}{dt}\rightarrow\infty$, and each $\Phi_n$ is a standing wave.

\section{P\"oschl-Teller Method}
\renewcommand{\theequation}{B.\arabic{equation}} \setcounter{equation}{0}

Since the problem in this work is mathematically equivalent to an infinite potential well, the P\"oschl-Teller method can also be used to approximately solve for the quasi-normal mode frequencies of this problem.
For linear ordinary differential equations,
 \bqn
 \lb{PT1}
\frac{d^2\Phi}{dy^2}+(\omega^2-V_{PT})\Phi=0.
 \eqn
and the PT potential requires
 \bqn
 \lb{PT2}
V_{PT}(y)=\alpha_o^2\left[\frac{K(K-1)}{\sin^2(\alpha_o(y-y_o))}+\frac{\lambda(\lambda-1)}{\cos^2(\alpha_o(y-y_o))}\right]\nb\\
 \eqn
the solution to the eigenvalue equation is given by
 \bqn
 \lb{PT3}
\omega_n&=&\alpha_o(K+\lambda+2n),\nb\\
\Phi_n&=&\frac{1}{\sqrt{C_o}}\sin^K(\alpha_o(y-y_o))\cos^\lambda(\alpha_o(y-y_o))\nb\\
&\times&{}_2F_1(-n,K+\lambda+n;K+\frac{1}{2};\sin^2(\alpha_o(y-y_o))),\nb\\
C_o&=&\frac{n!}{2\alpha_o}\left[\frac{\Gamma(K+1/2)}{\Gamma(K+1/2+n)}\right]^2\frac{\Gamma(n+1/2+K)\Gamma(n+1/2+\lambda)}{\Gamma(2n+1+K+\lambda)}\prod_{i=n}^{2n-1}(i+K+\lambda)
 \eqn
where we choose $C_o$ so that $\int_{y_0}^{y_0+\frac{\pi}{2\alpha_o}}\Phi^2(y)dy=1$. In this solution, $\alpha_o=\frac{\pi}{2y_\text{Max}}$ and we can choose $K$ and $\lambda$ to fit the axial gravitational potentials of this spacetime. we show the results in Table I

\begin{table}[htbp]
\caption{\label{Table1} Axial gravitational normal mode frequency $w$ obtained by the Shooting Method (SM) and P\"oschl-Teller Method(PT) with $\rho=1.5$ and $y\in[0,y_\text{Max}=2.90426]$. Here, we use potentials of P\"oschl-Teller as follows: $K=3.00001$, $\lambda=1.00036$ as $L=2$, $K=4.00003$, $\lambda=1.00073$ as $L=3$ and $K=5.00004$, $\lambda=1.00122$ as $L=4$.
}
\centering
\begin{tabular}{c c c c c c c}
         \hline\hline
 &$n=0$(SM)~~~~&$n=0$(PT)~~~~&$n=1$(SM)~~~~&$n=1$(PT)~~~~&$n=2$(SM)~~~~&$n=2$(PT)~~~~\\
        \hline
$L=2$~~~~ &$2.23507$~~~~&$2.16364$~~~~&$3.29516$~~~~&$3.24536$~~~~&$4.366$~~~~&$4.32708$~~~~\\
$L=3$~~~~ &$2.83101$~~~~&$2.70471$~~~~&$3.89405$~~~~&$3.78643$~~~~&$4.95426$~~~~&$4.86815$~~~~\\
$L=4$~~~~ &$3.41921$~~~~&$3.24584$~~~~&$4.4904$~~~~&$4.32756$~~~~&$5.54809$~~~~&$5.40928$~~~~\\
\hline
\hline
\end{tabular}
\end{table}

From the above results, it can be seen that, unfortunately, the results of the P\"oschl-Teller method are not very accurate. This is because the potential function in this case is discontinuous, and the right boundary of the potential function is very steep, making it difficult to achieve a very good fit using the two parameters $K$ and $\lambda$. Nevertheless, because the P\"oschl-Teller method is very intuitive and simple, we can use it to estimate the eigenfrequencies.

\begin{acknowledgments}
The authors thank anonymous referees for their useful and physically relevant comments. This work is supported by the Brazilian agencies FAPESQ-PB and CNPq. A.R.Q. acknowledges support from CNPq under process number 310533/2022-8.

\end{acknowledgments}

\end{document}